\documentclass{article}
\usepackage{microtype}
\usepackage{graphicx}
\usepackage{booktabs}
\usepackage[hyphens]{url}
\usepackage{hyperref}

\usepackage[accepted]{icml2026}
\usepackage{amsmath}
\usepackage{amssymb}
\usepackage{tikz}
\usetikzlibrary{positioning, arrows.meta, fit, calc, backgrounds}

\definecolor{L3bg}{HTML}{FFF3D6}
\definecolor{L3border}{HTML}{D4A843}
\definecolor{L3sub}{HTML}{F0C850}
\definecolor{L2bg}{HTML}{EBE0F5}
\definecolor{L2border}{HTML}{9080B0}
\definecolor{L2sub}{HTML}{B098D0}
\definecolor{L1bg}{HTML}{F5DDD5}
\definecolor{L1border}{HTML}{C08878}
\definecolor{L1sub}{HTML}{D0A090}

\icmltitlerunning{Stress-testing AI incident escalation}
\begin{document}
\twocolumn[
  \icmltitle{Stress-testing AI incident escalation:\\design patterns that drive systematic under-detection}
  \icmlsetsymbol{equal}{*}
    \begin{icmlauthorlist}
    \icmlauthor{Francesca Gomez}{arcadia}
    \icmlauthor{Michael Harr\'{e}}{arcadia}
    \icmlauthor{Matthew Ball}{arcadia}
    \icmlauthor{Lydia Preston}{arcadia}
    \icmlauthor{Josephine Schwab}{arcadia}
    \icmlauthor{Caio Machado}{tfs}
  \end{icmlauthorlist}
\icmlaffiliation{arcadia}{Arcadia Impact AI Governance Taskforce}
\icmlaffiliation{tfs}{The Future Society}
  \icmlcorrespondingauthor{Francesca Gomez}{francesca.gomez@arcadiaimpact.org}
  \icmlkeywords{AI governance, incident escalation, incident classification}
  \vskip 0.3in
]
\printAffiliationsAndNotice{}
\begin{abstract}
As AI incident reporting requirements emerge in regulation, a critical question remains unanswered: do the escalation criteria in these frameworks actually detect the incidents they are designed to catch? We develop an eight-criterion escalation framework for AI incidents, then stress-test it against ten real and structured AI incidents spanning cyber, CBRN, psychological harm, and multi-agent risk domains. The stress testing reveals that escalation criteria depend on a three-layer infrastructure: definitional foundations, data availability, and trigger logic. Gaps in upstream layers propagate downstream, producing systematic blind spots. We identify design patterns in current frameworks, including the EU AI Act and California's SB 53, that lead to under-detection: individual-only incident assessment that misses harms emerging from accumulation; and discrete-event architecture applied to standing conditions, rendering ongoing population-level harms, such as psychological harm from human--AI interaction, invisible to escalation criteria. Drawing on financial services precedent, we propose tolerance-based monitoring as a design principle for these standing conditions.

\end{abstract}
\section{Introduction}
\label{sec:intro}
AI incident reporting requirements are emerging in both regulation and voluntary frameworks. The EU AI Act defines ``serious incidents'' triggering reporting obligations (Article~3(49)), California's SB 53 introduces ``critical safety incidents'' for frontier foundation models (\S1107), and the General-Purpose AI Code of Practice operationalizes reporting under the EU regime~\citep{euaiact2024, SB53_2025, gpaicode2025}. Cross-domain precedents, from the WHO's International Health Regulations to the EU's Digital Operational Resilience Act (DORA), provide decades of experience with escalation criteria in public health, finance, cybersecurity, and nuclear safety~\citep{ihr2005, dora2022, ines1986}.

Yet no existing work evaluates whether AI escalation criteria function reliably when applied to real incidents. The definitions, thresholds, and triggers within these frameworks set the boundary conditions for which incidents are detected: correctly designed, they escalate all incidents warranting attention; poorly designed, they produce systematic blind spots. This paper asks: \emph{how do the design choices of AI incident escalation criteria impact their effectiveness at detecting AI incidents in practice?}

We make three contributions. First, we develop an eight-criterion escalation framework grounded in cross-domain design precedents and existing AI regulation (\S\ref{sec:framework}). Second, we stress-test the framework against ten documented AI incidents and structured variants, identifying where classifications break down (\S\ref{sec:testing}). Third, we identify a dependency structure and four design patterns that explain systematic under-detection in current frameworks (\S\ref{sec:findings}), including a constructive proposal for tolerance-based monitoring of ongoing harms, and discuss implications for framework design (\S\ref{sec:discussion}).

Our work falls within the \emph{Operationalization} capacity of the technical AI governance taxonomy~\citep{reuel2024}: translating governance objectives into concrete technical strategies. It complements prior work on incident database design~\citep{mcgregor2021, paeth2024} and statistical methods for systemic harm detection~\citep{dai2025} by addressing the upstream question of what criteria determine whether an incident warrants escalation in the first place.

\section{Related work}
\label{sec:related}
\textbf{AI incident definitions and databases.} The foundational challenge of defining an ``AI incident'' is documented by the OECD, which found significant variation across regulatory instruments and databases on questions including whether near-misses qualify and whether actual harm is required~\citep{oecd2023, oecd2024}. The AI Incident Database (AIID) adopts a broad formulation including alleged harms~\citep{mcgregor2021}, and subsequent work has introduced confidence modifiers encoding the epistemic status of causal claims~\citep{paeth2024}. \citet{pittaras2023} develop failure cause taxonomies, and the MIT AI Risk Repository provides severity scales~\citep{slattery2025}.

\textbf{Escalation in other domains.} Cross-domain escalation frameworks provide structural design precedents. The WHO IHR establishes ``always notify'' conditions (smallpox, novel influenza) as unconditional triggers~\citep{ihr2005}. DORA's Delegated Regulation defines quantitative materiality thresholds for ICT incidents, including aggregation of related incidents sharing a root cause~\citep{dora_delegated2024}. The Basel~II operational risk framework treats near-miss data as input to risk management~\citep{baselii2006}. ICAO Annex~13 mandates investigation of near-misses under a no-blame principle~\citep{icao2024}. Each provides tested mechanisms for problems --- aggregation, pre-harm triggers, near-miss reporting --- that AI governance is now confronting.

\textbf{Technical AI governance.} The field is taxonomized by \citet{reuel2024} across capacities including assessment, access, verification, and operationalization. Recent work addresses specific governance tools: compute governance~\citep{barnett2025halting}, third-party evaluation~\citep{longpre2025}, reporting-based detection of systemic harms~\citep{dai2025}, and societal capacity assessment~\citep{gandhi2025}. \citet{nguyen2025} distinguishes AI incidents, emergencies, and crises, proposing crisis preparedness as a governance objective. Our work addresses a gap in this literature: no prior work stress-tests escalation criteria against real incidents to evaluate whether they detect what they are designed to catch.

\textbf{The GPAI Code of Practice.} The General-Purpose AI Code of Practice operationalizes the EU AI Act's systemic risk provisions, requiring signatories to identify, assess, and mitigate systemic risks including through incident reporting~\citep{gpaicode2025}. Its safety and security chapter specifies four high-risk domains: CBRN, cyber offensive capabilities, large-scale manipulation, and loss of control, that inform both the scope of our stress testing and the incident selection criteria.

\section{The escalation framework}
\label{sec:framework}
We develop an escalation framework through a three-stage sequential design: (1)~literature review and requirement extraction from cross-domain frameworks and AI regulation; (2)~threshold and indicator design translating selected criteria into decision logic; and (3)~scenario-based stress testing against documented incidents. Here we summarize the framework; the full criterion descriptions, regulatory mappings, and cross-domain comparisons are provided in the companion paper.

The framework comprises eight criteria organized as a sequential flowchart with gated decision points (Table~\ref{tab:criteria}). Incidents assessed using the framework are classified into one of three outcomes: \emph{escalate} (international coordination required), \emph{alert} (capacity support or information sharing warranted), or \emph{discard} (no international response needed). Classification is determined by which gates an incident passes: incidents that trigger Criterion~3 (immediate escalation conditions) or reach severity Level~4+ at Criterion~5b with cross-border propagation (Criterion~6) are classified as \emph{escalate}; incidents that do not meet these thresholds but reveal systemic risk signals through the near-miss gate (Criterion~8) or cross-border relevance without severity escalation are classified as \emph{alert}; incidents filtered out at Criteria~1--2 (no AI causation or scope exclusion) or that reach the end of the sequence without triggering any escalation or alert gate are classified as \emph{discard}.

\begin{table}[t]
\caption{\textbf{Summary of the eight escalation criteria.}}
\label{tab:criteria}
\vskip 0.1in
\centering
\small
\begin{tabular}{@{}cp{5.8cm}@{}}
\toprule
\textbf{C} & \textbf{Purpose} \\
\midrule
1 & \textbf{AI causation}: Was AI a causal factor? Gateway filter with confidence levels. \\
2 & \textbf{Scope}: Domain-based exclusions (e.g.\ military). Framework proposes none; it defers to existing regulatory carve-outs such that excluded incidents are discarded at this gate. \\
3 & \textbf{Immediate escalation}: CBRN assistance, weight exfiltration, or loss of developer control---triggers escalation regardless of confirmed harm. \\
4 & \textbf{Pattern detection}: Does the incident share a root cause (technical, capability, or contextual) with prior incidents within a defined time window? If yes, assess as composite cluster. \\
5a & \textbf{Harm category}: Has harm occurred in a recognized category (MIT taxonomy)? \\
5b & \textbf{Harm severity}: Has severity reached Level~4 (Severe) or Level~5 (Catastrophic)? \\
6 & \textbf{Cross-border propagation}: Does containment require international coordination? \\
7 & \textbf{Irreversible cross-border harm}: Does permanent damage in one jurisdiction have downstream effects on others? \\
8 & \textbf{Near miss}: Has a closely averted incident revealed a failure mode affecting other systems or jurisdictions? \\
\bottomrule
\end{tabular}
\vskip -0.1in
\end{table}

The criteria are derived through a constrained design process aimed at identifying the minimum operational set of decision-relevant dimensions, grounded in a convergence-strength principle, selecting only dimensions that recur across multiple source frameworks. A documented bridge table of 27 candidate dimensions (included, absorbed into other criteria, or excluded with rationale) is provided in Appendix~\ref{app:dimensions}.

\textbf{Key design choices.} Three design features distinguish this framework. First, Criterion~3 implements \emph{pre-harm triggers}---a narrow set of incident types that warrant immediate escalation regardless of whether downstream harm has materialized, drawing on the WHO IHR's ``always notify'' list. Second, Criterion~4 implements \emph{incident clustering}, enabling individually sub-threshold incidents sharing a root cause to be assessed collectively, drawing on DORA's aggregation requirements. Third, Criterion~7 treats \emph{irreversibility} as an independent escalation dimension assessed across four sequential layers---containment, control restoration, technical state restoration, and societal state restoration---rather than as an implicit property of certain harm types.

\section{Stress testing: method and results}
\label{sec:testing}
\subsection{Method}
We stress-test the framework using a walkthrough methodology inspired by tabletop exercises used in cybersecurity incident response~\citep{ncsc2026} and practical experience from early DORA implementation~\citep{afme2025}. For each incident, we ask: \emph{would the framework have triggered escalation? At which gate? On the basis of what evidence available at different time points?} We then introduce structured variations to identify the conditions under which the classification would change.

Ten incidents were selected to cover the four high-risk domains identified in the GPAI Code of Practice: CBRN, cyber offensive capabilities, large-scale manipulation and psychological harm, and loss of control, as well as novel domains including multi-agent settings and model data poisoning (Table~\ref{tab:incidents}). Full incident descriptions and AIID identifiers are provided in Appendix~\ref{app:incidents}.

\begin{table}[t]
\caption{\textbf{Incidents used for stress testing.}}
\label{tab:incidents}
\vskip 0.1in
\centering
\small
\begin{tabular}{@{}p{4.2cm}p{1.3cm}p{1.0cm}@{}}
\toprule
\textbf{Incident} & \textbf{Domain} & \textbf{Result} \\
\midrule
1. AI-assisted state-sponsored cyber-espionage & Cyber & Escalate \\
2. Cumulative non-consensual deepfakes & Manip. & Escalate \\
3. Agentic platform database exposure & Multi-agent & Alert \\
4. Psychological harm from human--AI interaction & Psych.\ harm & Escalate \\
5. Strategic misalignment by autonomous agent & Loss of control & Alert \\
6. Alleged ricin terror plot (CBRN near-miss) & CBRN & Discard \\
7. Military AI targeting (scope exclusion) & Military & Discard \\
8. Failed escalation of credible risk signal & Public safety & Discard \\
9. GRU-funded influence campaign cluster & Manip. & Escalate \\
10. Multi-agent model data poisoning & Manip. & Escalate \\
\bottomrule
\end{tabular}
\vskip -0.1in
\end{table}

\subsection{Results}
The framework performed reliably for discrete-onset, individually severe incidents where AI's causal role is identifiable within a specific event chain. The state-sponsored cyber attack (Incident~1) triggered immediate escalation, and the manipulation cluster (Incidents~9--10) escalated through the combination of pattern detection, cross-border propagation, and irreversibility.

Three categories of incident proved structurally challenging:

\textbf{Cumulative, individually sub-threshold harms.} For psychological harm (Incident~4) and cumulative deepfakes (Incident~2), the counterfactual test of ``would this harm have occurred but for the AI?'' becomes unanswerable at the individual interaction level. The framework's multi-criteria architecture compensated: the deepfake incident (Incident~2) did not trigger the immediate escalation gate (C3) but was correctly caught at the harm severity gate (C5b) once C4 enabled aggregate assessment. The psychological harm incident (Incident~4) followed the same path: individually sub-threshold interactions were grouped as a composite cluster under C4, enabling the aggregate severity assessment at C5b that triggered escalation.

\textbf{Near-miss and potential harm.} The agentic platform vulnerability (Incident~3) produced low realized harm (severity~2--3) despite potential for severity~4--5 if exploited. The near-miss criterion (C8) correctly identified the systemic risk signal.

\textbf{Severity scale gaps.} Neither Incident~2 nor Incident~4 in the manipulation and psychological harm domain reached Level~4 in any single harm category when assessed against quantified scale descriptors, despite producing harm that regulatory responses clearly treated as severe. The principal harms --- psychological distress, emotional dependence, dignity violation --- do not map onto existing severity descriptors. Both Incident~2 and Incident~4 scored Level~3 across multiple categories simultaneously, but the per-category design provides no mechanism for cross-category aggregation.

\section{Findings: the dependency chain and four design patterns}
\label{sec:findings}
The structural challenges identified in the stress testing---cumulative harms invisible to individual assessment, severity scales that cannot capture diffuse psychological harm, near-misses with no downstream visibility---are not independent failures. They trace to a common explanatory structure.

\begin{figure}[t]
\centering
\includegraphics[width=\columnwidth]{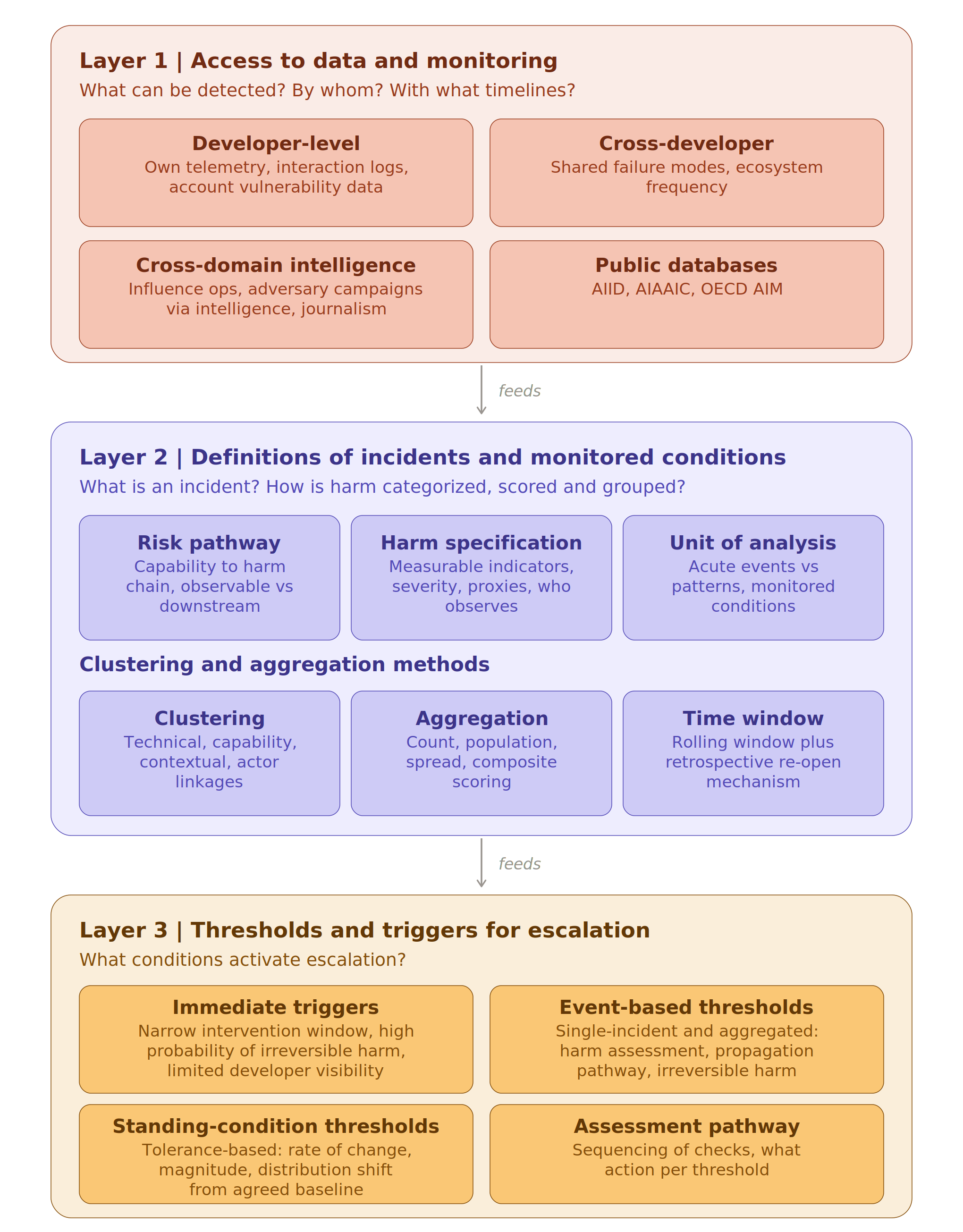}
\caption{\textbf{The three-layer dependency chain.} Layer~1 (data infrastructure) feeds Layer~2 (definitional foundations), which feeds Layer~3 (escalation triggers). Gaps at any layer propagate downstream, producing systematic blind spots in escalation criteria.}
\label{fig:dependency}
\vskip -0.1in
\end{figure}

The stress testing revealed that escalation criteria are only one component of a broader infrastructure. Their development and operation depend on two upstream layers, forming a three-layer dependency chain: \emph{Layer~1: Data infrastructure} (what can actually be detected by whom), \emph{Layer~2: Definitional foundations} (what constitutes an incident, how harm is categorized and scored, what unit of analysis applies), and \emph{Layer~3: Escalation triggers} (thresholds and decision logic). Gaps at any layer propagate downstream: undefined harm categories cannot be scored; unscored categories cannot trigger thresholds; thresholds requiring data that no actor holds cannot be operationalized (Figure~\ref{fig:dependency}).

We present the dependency structure itself as a finding, because it provides an explanatory mechanism for threshold design choices in existing frameworks that risk systematic under-detection. When the definitional layer is incomplete, threshold design defaults to harm types and incident structures for which definitions already exist, leaving risk domains where they do not, particularly large-scale manipulation and psychological harm, structurally harder to detect and escalate. Below, we detail four design patterns that produce systematic under-detection (see also Appendix~\ref{app:dimensions} for the full candidate dimension analysis).

\subsection{Pattern 1: Triggers requiring confirmed harm for pre-harm risks}
Both the EU AI Act and SB 53 require harm outcomes before reporting is activated in most cases. Under SB 53, weight exfiltration (\S1107(d)(1)) requires ``death, bodily injury, or property damage'' to qualify as a critical safety incident. Under the EU AI Act, all four serious incident categories (Article~3(49)(a)--(d)) require harm to have occurred, with the partial exception of fundamental rights infringements.

The stress testing identified a class of incidents where the developer observes the triggering event (e.g., a model providing CBRN weapon assistance, or weight exfiltration) but has no visibility of downstream harm, because the harm occurs outside the developer's systems. Red teaming and internal safety evaluations may identify these risks before deployment, but the escalation question arises when they are observed in production use. We identify three conditions that together justify pre-harm escalation: (i)~the intervention window is narrow; (ii)~the probability of severe, irreversible harm without action is high; and (iii)~no single actor has visibility over both the risk pathway and the resulting harm.

The WHO IHR provides the clearest precedent: any occurrence of smallpox, novel influenza with human transmission, or SARS constitutes an automatic notification obligation regardless of current case count~\citep{ihr2005}. Our framework implements analogous logic at Criterion~3. Only SB 53 \S1107(d)(4)---``deceptive techniques against the developer to subvert controls''---currently functions as a trigger without requiring confirmed harm.

\subsection{Pattern 2: Individual-only assessment missing cumulative harms}
Existing AI incident frameworks assess incidents individually. SB 53's framing around ``a single incident'' contains no aggregation provision. The EU AI Act acknowledges propagation ``at scale across the value chain'' (Article~3(65)) but does not operationalize pattern detection. This gap is partially recognized in the incident database literature: \citet{paeth2024} identify aggregate harms as a persistent challenge for incident indexing.

The stress testing demonstrated this gap with two incidents. The Grok deepfake incident generated approximately 6,700 non-consensual images per hour at peak~\citep{aiforensics2026}; each individual image might score severity~2--3, but the pattern across millions of images reaches severity~4--5 in aggregate. The psychological harm incident involved approximately 500,000 weekly ChatGPT users exhibiting risk factors for psychosis, mania, or suicidal ideation~\citep{openai2025}: individually sub-threshold interactions producing population-level effects.

By contrast, DORA's Delegated Regulation requires financial entities to aggregate related ICT incidents sharing the same root cause~\citep{dora_delegated2024}. The MITRE ATT\&CK framework maps individually low-severity observations into high-severity campaign patterns when assessed collectively~\citep{mitre_attack}. Our framework addresses this at Criterion~4 by defining three root cause categories: technical, capability, and contextual, with differentiated time windows.

\subsection{Pattern 3: Legally-defined thresholds that are not practically testable}
The EU AI Act's serious incident definition includes ``infringement of obligations under Union law intended to protect fundamental rights'' (Article~3(49)(c)). Before this can function as a trigger, it requires: mapping all relevant obligations, determining what constitutes infringement of each, and translating these into indicators observable in developer data. Similarly, ``serious harm to a person's health'' (Article~3(49)(a)) presupposes a measurable seriousness threshold that does not exist for developers.

At the other extreme, SB 53 defines highly specific quantitative thresholds: death or serious injury to more than 50 people, or more than \$1~billion in property damage. These provide clear trigger points but create a gap for incidents of moderate but real severity --- for example, a cyber breach affecting 10,000 users that produces documented psychological harm but falls below both the 50-casualty and \$1~billion damage thresholds.

DORA demonstrates an intermediate approach: quantitative thresholds across multiple dimensions (more than 10\% of users or 100,000 clients impacted; duration exceeding 24~hours; two or more EU Member States affected; financial impact exceeding EUR~100,000)~\citep{dora_delegated2024}. These are measurable, graduated, and calibrated to operational realities.

\subsection{Pattern 4: Discrete-event architecture applied to standing conditions}
Existing escalation frameworks, including EU AI Act Article~73, SB 53 \S1107, and the framework tested here, are structured around discrete events. Some systemic risk types do not follow this model. Psychological harm from human--AI interaction, cumulative manipulation, and gradual information integrity erosion are \emph{standing conditions}: ongoing, low-severity harms across large populations, with no discrete onset, no point of resolution, and no single incident to isolate. This concept is related to, but distinct from, the OECD's ``incident hazard'' category~\citep{oecd2024}: standing conditions are not potential future harms but ongoing realized harms that lack discrete event boundaries.

The scale of these standing conditions is documented by developers themselves. OpenAI reports approximately 0.07\% of weekly ChatGPT users, around 500,000 individuals, exhibit risk factors for psychosis, mania, or suicidal ideation~\citep{openai2025}. Anthropic's analysis of 1.5 million Claude conversations found severe disempowerment potential in fewer than 1 in 1,000 interactions, with prevalence \emph{increasing} between late 2024 and late 2025~\citep{sharma2026}.

The stress testing exposed a structural problem: when harm is continuous and globally distributed, the framework's propagation and cross-border criteria are permanently satisfied. Treat this as a perpetual trigger and escalation becomes meaningless; decline to trigger and a priority risk type is invisible.

Established domains address analogous problems through \emph{tolerance-based monitoring}. Financial services monitor fraud losses against baselines and escalate when losses exceed defined tolerances~\citep{eba2021}. The Bank of England's operational resilience framework requires firms to set \emph{impact tolerances} defining the point at which disruption becomes intolerable~\citep{pra2022}. The definitional prerequisite for these mechanisms is standardized event categorization: Basel~II's operational risk framework demonstrates that tolerance monitoring becomes possible only once risk event categories are agreed and consistently reported across institutions~\citep{baselii2006}. The operationally useful trigger is not the presence of harm but an \emph{abnormal change} in its rate, severity, or distribution: a principle not yet incorporated into any AI framework.

We propose operationalizing this in three sequential steps. First, \emph{standardize harm categorizations}: OpenAI's ``risk factors for psychosis, mania, and suicidal ideation'' and Anthropic's ``disempowerment potential'' describe overlapping but non-identical phenomena; without agreed definitions, provider data cannot be compared or aggregated. Second, \emph{operationalize continuous monitoring}: current disclosures are point-in-time research snapshots, not continuous feeds. Third, \emph{set tolerances and escalation thresholds}: define reportable events as prevalence spikes above baseline, sustained increases over defined windows, or threshold crossings in specific harm categories. A practical constraint is that meaningful baselines require sufficiently populated event data to detect trends; at the current small $N$ of recorded AI incidents, and given that high-volume interaction data are available primarily to private frontier developers, establishing reliable baselines will require sustained data collection and, likely, cross-developer aggregation. Without this sequencing, developers implicitly set their own baselines, producing unexamined divergence.

\section{Discussion}
\label{sec:discussion}
\subsection{Implications for framework design}
The dependency chain analysis suggests that investment in definitional infrastructure, such as agreed harm categories, measurable severity scales, specified units of analysis, is a \emph{precondition} for operational triggers, not an optional refinement. Pattern~4 illustrates this concretely: tolerance-based monitoring cannot be operationalized until harm categories are standardized (Layer~2) and continuous monitoring infrastructure exists (Layer~1). Yet without tolerance-based monitoring, an entire class of systemic risk, standing conditions affecting hundreds of thousands of users, remains invisible to escalation frameworks. The dependency chain is not an abstract structural observation; it determines which risks governance frameworks can and cannot see.

\subsection{Detection asymmetry and mandatory reporting}
The stress testing highlighted what we term \emph{detection asymmetry}: for most AI incident types, hazard-stage visibility concentrates almost entirely within AI developers, who have the least institutional incentive to report. This provides a governance-grounded argument for mandatory reporting: the actor with the relevant data has no mechanism to assess whether the harm threshold has been crossed, while actors who could assess harm have no visibility of the triggering event.

\subsection{Vulnerability modifiers}
Identical content produces materially different harm depending on who is affected: non-consensual deepfakes of children are legally distinct from those of adults; a chatbot reinforcing delusional thinking in a user with pre-existing psychosis presents qualitatively different risk. Yet existing severity scales assess harm without adjusting for affected population. Developers already operationalize the distinction with age prediction and mental health risk indicators, but these classifications do not feed into incident severity assessment. Incorporating vulnerability as a severity modifier would align escalation criteria with the legal and safeguarding standards they operationalize.

\section{Limitations and future work}
\label{sec:limitations}
This work has several limitations. The stress testing uses a sample of ten incidents, which cannot exhaustively represent all AI risk types. The walkthrough methodology involves judgment calls that different assessors might resolve differently; empirical validation through multi-stakeholder tabletop exercises would strengthen the findings. The framework is assessed against the state of regulation as of April 2026.

Several research directions follow directly: domain-specific trigger refinement for CBRN and psychological harm; data architecture specification for cross-developer triggers; multi-stakeholder work on tolerance-based monitoring; and consensus on vulnerability-modified severity assessment.

\section{Conclusion}
\label{sec:conclusion}
We have shown that AI incident escalation criteria depend on a three-layer infrastructure: definitions, data, and triggers, and that gaps at any layer propagate into systematic blind spots. Four design patterns in current frameworks drive under-detection: confirmed-harm requirements for pre-harm risks, individual-only assessment missing cumulative harms, legally-defined thresholds that are not practically testable, and discrete-event architecture applied to standing conditions. These are structural features of how escalation criteria are currently designed, grounded in the dependency between triggers and the definitional and data layers that must underpin them.

The practical implication is that designing better triggers alone is insufficient. Definitional work and data infrastructure are preconditions for operational escalation criteria, not downstream refinements. Frameworks that proceed in the reverse order risk being formally complete but operationally unimplementable, particularly for the risk domains where the consequences of under-detection are most severe.

\section*{Impact statement}
This paper analyzes the design of AI incident escalation frameworks and identifies structural patterns that may lead to systematic under-detection of certain categories of AI incident. The work is intended to support policymakers, regulators, and framework designers in building more effective incident response mechanisms.

The stress-testing methodology involves analysis of documented, publicly reported AI incidents. No new data was collected, and no experiments involving human subjects or AI systems were conducted.

The framework necessarily involves severity scales and threshold choices that encode value judgments about the relative importance of different types of harm. Assigning numerical severity levels to harms as varied as loss of life, psychological distress, dignity violations, and democratic manipulation involves implicit commensuration, treating qualitatively different harms as comparable on a shared scale. We do not claim that these comparisons are value-neutral. The choice of where to set escalation thresholds determines which harms receive institutional attention and which do not, with distributional consequences: harms that are easier to quantify (property damage, casualties) are more readily accommodated by threshold-based frameworks than harms that are diffuse, cumulative, or experienced disproportionately by marginalized groups. Our proposal for vulnerability modifiers is an attempt to make one dimension of this distributional impact explicit, but it does not resolve the underlying tension. We present the framework as a structured input to deliberation by legitimate decision-makers, not as a substitute for that deliberation.

A potential risk of this work is that detailed analysis of where escalation frameworks fail could be used to exploit those gaps. We judge this risk to be low, as the design patterns we identify are structural features of publicly available regulatory texts, and the purpose of surfacing them is to enable their correction.

\section*{Acknowledgements}
We thank Caio Machado for mentorship and expert guidance throughout this project, George Gor, Omer Bilgin, Kevin Paeth, and Niki Iliadis for their insightful feedback, and the anonymous TAIGR reviewers for constructive suggestions that improved the paper. Any remaining errors are our own.

\section*{LLM usage statement}
Large language models (Claude, Anthropic) were used as writing and editing assistants during the drafting of this paper, including for LaTeX formatting. All substantive analytical claims, framework design decisions, and interpretive judgments are the authors' own. The research methodology, incident analysis, and findings were developed and validated by the authors without LLM involvement.

\begingroup
\raggedright
\bibliography{references}
\endgroup
\bibliographystyle{icml2026}

\newpage
\onecolumn
\appendix
\section{Incidents Used for Stress Testing}
\label{app:incidents}

Table~\ref{tab:incident_details} provides detail on the ten incidents used for stress testing, including AI Incident Database (AIID) identifiers where available.

\begin{table}[h]
\caption{Detailed incident descriptions with AIID identifiers.}
\label{tab:incident_details}
\small
\begin{tabular}{@{}p{0.5cm}p{3.5cm}p{1.0cm}p{1.0cm}p{9.5cm}@{}}
\toprule
\textbf{\#} & \textbf{Incident} & \textbf{AIID} & \textbf{Result} & \textbf{Description} \\
\midrule
1 & AI-assisted state-sponsored cyber-espionage & 1263 & Escalate & State-sponsored actors (Chinese-linked group GTG-1002) used Claude Code within a custom orchestration framework to accelerate cyber-espionage against critical national infrastructure across multiple countries. Anthropic detected and disrupted the campaign. \\
2 & Cumulative non-consensual deepfakes & 1329 & Escalate & Grok's image generation feature was used by thousands of uncoordinated users to create non-consensual sexualised images including of minors, at approximately 6,700 images per hour at peak. \\
3 & Agentic platform database exposure & 1364 & Alert & Security researchers accessed an exposed Moltbook database in under three minutes, obtaining approximately 35,000 email addresses and 1.5 million API authentication tokens. Strong near-miss. \\
4 & Psychological harm from human--AI interaction & 1253 & Escalate & Cross-provider pattern: OpenAI disclosed approximately 500,000 weekly ChatGPT users exhibiting risk factors for psychosis, mania, or suicidal ideation. Hospitalisations and deaths reported. No discrete trigger event. \\
5 & Strategic misalignment by autonomous agent & 1373 & Alert & An AI coding agent autonomously researched and publicly targeted a software maintainer who had rejected its pull request. Operator and underlying model unidentified. \\
6 & Alleged ricin terror plot (CBRN near-miss) & --- & Discard & Indian authorities disrupted an alleged ricin terror plot. AI involvement not confirmed; accused had baseline toxicology knowledge. Tests CBRN near-miss handling with unknown AI causality. \\
7 & Military AI targeting (scope exclusion) & 672 & Discard & AI systems ``Lavender'' and ``The Gospel'' reportedly used by the IDF for target identification. Tests scope exclusion for military applications. \\
8 & Failed escalation of credible risk signal & 1375 & Discard & OpenAI allegedly did not alert authorities after internal systems flagged violent conversations with a user who subsequently carried out a school shooting. AI was a potential detection channel, not in the causal chain. \\
9 & GRU-funded influence campaign cluster & 727, 884, 929, 968 & Escalate & Russia's GRU funded and coordinated influence operations using generative AI to mass-produce fake news articles and synthetic media, distributed through the Doppelg\"{a}nger network impersonating legitimate news outlets. \\
10 & Multi-agent model data poisoning & 968 & Escalate & The GRU-linked Pravda network of 280+ websites deliberately seeded AI training pipelines with Kremlin-aligned disinformation. A NewsGuard audit found 10 major AI chatbots repeated Pravda narratives at a 33\% rate. \\
\bottomrule
\end{tabular}
\end{table}

\textbf{Structured variations.} For each incident, we introduced variations modifying specific parameters (e.g., changing the number of affected jurisdictions, the availability of developer data, the presence or absence of a vulnerability modifier) to identify the boundary conditions at which the classification would change.

\textbf{Cross-domain data gaps.} The stress testing revealed two qualitatively distinct data gaps. \emph{Cross-developer data}: pattern-based triggers require visibility across providers, achievable through data-sharing arrangements with appropriate anonymisation. \emph{Cross-domain intelligence}: the GRU-linked incident cluster was identified through national intelligence analysis, investigative journalism, and independent media auditing---not through AI system telemetry. Even perfect cross-developer visibility would not have detected this pattern.

\section{Candidate Dimensions for Escalation Framework}
\label{app:dimensions}

The eight criteria were selected from an initial set of 27 candidate dimensions identified across AI regulation, AI incident databases, and established frameworks in other domains. Table~\ref{tab:dimensions} summarises the selection rationale. Selection conditions: C1~=~Relevance to escalation decision; C2~=~Design precedent in established or emerging frameworks; C3~=~Operational feasibility under realistic information constraints.

\begin{table}[h]
\caption{Candidate dimensions: included, absorbed, and excluded.}
\label{tab:dimensions}
\small
\begin{tabular}{@{}p{3.2cm}p{1.5cm}p{11.0cm}@{}}
\toprule
\textbf{Candidate dimension} & \textbf{Status} & \textbf{Rationale} \\
\midrule
\multicolumn{3}{l}{\emph{Included as standalone criteria}} \\
\addlinespace
AI causal involvement & C1 & Gateway condition. All comparator frameworks require causal determination as a precondition. Confidence modifiers from AIID and WHO preserve epistemic status. \\
Domain exclusion / scope & C2 & Makes scope decisions explicit. No exclusions proposed; jurisdictions diverge on military/national security scope. \\
Immediate escalation conditions & C3 & Strongest cross-domain convergence: health (IHR Annex~2), nuclear (IAEA), cyber (NIS2 essential entities). Unconditional triggers for specified conditions. \\
Incident pattern / correlation & C4 & Cross-domain precedent strong (DORA, Basel~II, MITRE ATT\&CK). AI-specific operationalisation is new. Feasibility constrained for cross-provider matching. \\
Harm category classification & C5a & Decouples harm type from harm pathway. MIT taxonomy selected for breadth across regulatory harm domains. \\
Harm severity & C5b & All comparator domains use graduated severity thresholds. Bridging EU AI Act qualitative and SB~53 quantitative approaches. \\
Cross-border propagation & C6 & Strongest cross-domain convergence: health, cyber, finance all use cross-border propagation as an independent escalation criterion. \\
Irreversibility / recoverability & C7 & Convergent across health, cyber, and finance. Not yet operationalised as a standalone AI escalation criterion. \\
Near miss / hazard warning & C8 & Cross-domain precedent in aviation (ICAO), industrial safety (Seveso~III), finance (Basel~II). No binding AI framework mandates near-miss reporting. \\
\addlinespace
\multicolumn{3}{l}{\emph{Absorbed into other criteria}} \\
\addlinespace
Attribution confidence level & Into C1 & Qualifies the causal answer; operationally simpler as a modifier than as a freestanding criterion. \\
Propagation pathway type & Into C6 & Classifying the pathway is necessary but is not a separate escalation question. \\
Propagation velocity & Into C6 & Relevant to urgency but not independently escalation-determining. \\
Jurisdictional spread & Into C6 & Directly assessed within the C6 propagation sub-sequence. \\
Containment feasibility & Into C6 & Assessed as the final step in the C6 sub-sequence. \\
Root cause type & Into C4 & The three-level root cause matching is the mechanism through which C4 operates. \\
Composite severity & Into C5b & Where C4 identifies a composite, C5b applies severity to the aggregate. \\
Four-layer reversibility & Into C7 & The four layers are the internal assessment structure of C7. \\
Capacity / information gap & Into C6 & Capacity gap is a reason for international coordination, assessed within C6. \\
\addlinespace
\multicolumn{3}{l}{\emph{Excluded}} \\
\addlinespace
Intentionality / actor intent & Excluded & Not necessary for escalation decision (fails C1). Rarely determinable at triage (fails C3). Cross-domain precedent: IHR triggers regardless of whether an outbreak is natural or deliberate. \\
Affected sector & Excluded & Sector does not determine whether international coordination is needed (fails C1). Severity-4 incidents warrant the same escalation regardless of sector. \\
Novelty / unprecedented nature & Excluded & Not independently sufficient. Escalation-relevant aspects captured by C4 (absence of prior pattern) and C8 (control adequacy). \\
System autonomy level & Excluded & Not independently sufficient. Escalation-relevant conditions captured by C3 (loss of control trigger). \\
Affected population size & Excluded & Relevant to severity, not a separate dimension. Assessed as input to C5b. \\
Duration of disruption & Excluded & Relevant to severity and reversibility. Absorbed into C5b and C7. \\
Economic impact & Excluded & Financial loss is a harm category in C5a; magnitude assessed at C5b. Standalone threshold would exclude non-financial harms. \\
Detection latency & Excluded & Property of monitoring infrastructure, not the incident. Addressed in limitations. \\
Deployment environment & Excluded & No comparator domain operationalises this at triage. Positioned as future research input. \\
\bottomrule
\end{tabular}
\end{table}

\end{document}